\documentclass[twocolumn]{revtex4}
\usepackage{mathtools}
\usepackage{float}
\usepackage{subfigure}
\usepackage{paralist}
\usepackage{bm}
\usepackage{amsfonts}
\usepackage{url}
\usepackage{multirow}
\usepackage{color}
\usepackage{booktabs}
\usepackage{lipsum}
\usepackage[linesnumbered,ruled,vlined]{algorithm2e}
\usepackage{algpseudocode}
\usepackage{amsmath}
\usepackage{cleveref}
\usepackage{amssymb}

\begin{document}
	\title{Effective Model Integration Algorithm for Improving Link and Sign Prediction in Complex Networks}
	
	\author{Chuang Liu}
	\affiliation{Alibaba Research Center for Complexity Sciences, Hangzhou Normal University, Hangzhou, 311121, China}
	\author{Shimin Yu}
	\affiliation{Alibaba Research Center for Complexity Sciences, Hangzhou Normal University, Hangzhou, 311121, China}
	\author{Ying Huang}
	\affiliation{Institute of VR and Intelligent System,, Hangzhou Normal University, Hangzhou, 311121, China}
	\author{Zi-Ke Zhang}\email{zkz@zju.edu.cn}
	\affiliation{College of Media and International Culture, Zhejiang University, Hangzhou 310028, China}
	\affiliation{Alibaba Research Center for Complexity Sciences, Hongzhou Normal University, Hangzhou 311121, China}
	\affiliation{Center for Research on Zhejiang Digital Development and Governance, Hangzhou 310028}


	\begin{abstract}
Link and sign prediction in complex networks bring great help to decision-making and recommender systems, such as in predicting  potential relationships or relative status levels. Many previous studies focused on designing the special algorithms to perform either link prediction or sign prediction. In this work, we propose an effective model integration algorithm consisting of network embedding, network feature engineering, and an integrated classifier, which can perform the link and sign prediction in the same framework. Network embedding can accurately represent the characteristics of topological structures and cooperate with the powerful network feature engineering and integrated classifier can achieve better prediction. Experiments on several datasets show that the proposed model can achieve state-of-the-art or competitive performance for both link and sign prediction in spite of its generality. Interestingly, we find that using only very low network embedding dimension can generate high prediction performance, which can significantly reduce the computational overhead during training and prediction. This study offers a powerful methodology for multi-task prediction in complex networks.

\textbf{link prediction, sign prediction, complex network, deep learning} 		
\end{abstract}
	
	%
\maketitle

	\section{Introduction}\label{sec:introduction}
	
	\begin{figure*}[htb]
		\centerline{\includegraphics[width=13cm]{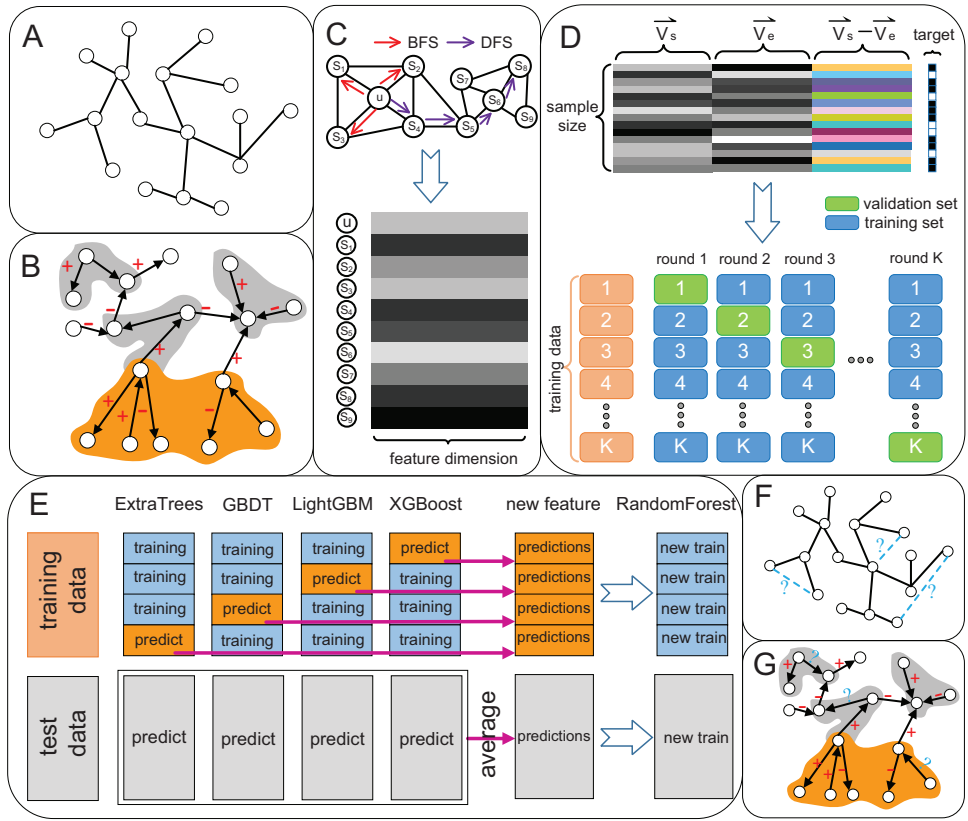}}
		\caption{Framework of integration model. (A) and (B) show an ordinary undirected unsigned network and a directed signed network as examples of link and sign prediction, where  orange and gray  respectively represent typical bridge edge structures and triangle edge structures, and  red +/- signs respectively represent  positive and negative relative status levels between two nodes. (C) Triangle and bridge structures are modeled simultaneously using  graph embedding (Node2Vec). (D) Strengthened features via  network characteristic engineering are used to build  training and validation sets in  K-fold cross-validation. (E) Integrated classifiers are trained to generate expanded features for (F)  link and (G) sign prediction.}
		\label{fig:framework}
	\end{figure*}
	
	A complex network is a set of connected nodes that describe different spatial, temporal, or logical relationships, such as social network \cite{borgatti2020science} and biological network \cite{liu2020pr}.  The studies of complex networks have become an important research filed, including the network prediction \cite{sun2020nc}, network spreading \cite{liu2020amc}, network controllability \cite{liu2011nature} and so on. In particular, the representation and prediction techniques of complex networks have been key issues in recent years.
	
	The network structure is the most basic information in the network research. However, the incompleteness of the network is a common problem in the real system for the limitation of the observation data. Therefore it is very important to predict the missing links to obtain a more reliable network structure \cite{lu2011pa}. In this work, we focus on the link and sign prediction of complex networks \cite{yuan2019graph,tuan2019fuzzy} to determine whether a link exists between two nodes, and if so, the relative status level. Applications of the link prediction  include predicting the interaction between proteins or genes in biological networks \cite{liu2020pcb}, whether people in a social network have a relationship and are friends or enemies \cite{wu2012diversity}, whether authors in academic networks will cooperate \cite{brandao2013using}, and even the scientific impact prediction \cite{Butun2020tc}. Recently, there have emerged signed networks having both positive and negative links. Relationships in networks, such as friends and enemies in social networks, are usually represented as "+" or "-." These unsigned and signed networks have different characteristics, indicating that the sign prediction is very different from the link prediction \cite{yaghi2020link}.
	
	The sparseness of network data such as real social networks brings great challenges to link and sign prediction \cite{martinez2016survey}. Many methods have been proposed, with three kinds of strategies: topological-based methods \cite{liben2007link,zhou2009predicting} that  use network-based similarity to determine the probabilities of potential links; model-based methods \cite{ghasemian2019evaluating,clauset2008hierarchical}, such as random block models and other community structure models; and embedding methods \cite{grover2016node2vec,cai2018comprehensive} that map a network to the latent space and predict whether links exist based on their proximity to latent vectors. Very recently, the embedding methods has attracted much attention in link prediction \cite{zhan2020susceptible,zhang2019degree}. The network embedding methods in the unsigned networks, including the \textit{DeepWalk} \cite{perozzi2014deepwalk}, \textit{Node2vec} \cite{grover2016node2vec}, \textit{LINE} \cite{tang2015line} and so on, aim to represent the linked nodes to be close to each other in embedding space. Typically, some methods are also proposed in the signed network embedding, including the \textit{Beside} \cite{chen2018bridge}, \textit{SIDE} \cite{kim2018side}, \textit{ASiNE} \cite{Lee2020sigir} and so on, with assuming that nodes with positive links tend to be close to each other while those with negative links tend to be far away from each other in embedding space. Many studies focus on network embedding with designing the better representation learning model to obtain a more accurate vector to represent the network node.
	
	However, it would be very challenging to design a perfect network embedding model. For example, the role of the negative links is not very clear in the signed network embedding \cite{Lee2020cikm}. The deep-learning based models, which show better representation ability in network embedding, would perform much more computations, especially for the large network data. And the embedding vectors can only efficient in some special prediction task \cite{Lee2020cikm}. Actually, obtaining the embedding vector of the network node can be considered as the feature extraction of the prediction task. In this work, we focus on the following prediction model after the network embedding, which is ignored in the previous studies. Using the node features by an acceptable embedding approach (eg. \textit{Node2vec} \cite{grover2016node2vec}), we aims to build an efficient classification model to perform the prediction task using the embedding vectors. In this way, we can generate very good prediction performance without designing a complicated network embedding model. In addition, many proposed methods can perform either link prediction or sign prediction. For example, the signed network embedding, such as \textit{Beside}, can only obtain the better performance on the sign prediction. And how to build a generalized model to perform link and sign prediction with multi-scene networks in the same framework \cite{Agrawal2019www} is also the motivation of this work.
	
	To solve the problem of performing link and sign prediction task in the same framework, we propose a  model integration algorithm to improve link prediction (considering sign prediction as a branch of link prediction), which we refer to as the integration model (or the mixed strategy \cite{Cao2020tc}). The model integration algorithm generates the classification results by combining many basic learners, and the prediction of the integration models is generally  more accurate than the basic learners \cite{agajanian2019,kaur2021}. The inputs of the integration model in this work are based on the network embedding features. After interpretable network feature engineering, we select some suitable machine learning classifiers as the basic learners, and train the final classifier with integrated learning. Experiments validate the model's performance at both link and sign prediction. The model can perform well on  network data where a large number of triangle and bridge structures coexist. It does not depend on the balance model,  is not limited to  a specific network substructure, and has good generalization ability. Our analysis of the impact of network embedding dimensions indicates that our model does not require high-dimensional embedding features, and low-dimensional embedding can achieve good results.
	
	The primary contributions of this work can be summarized as follows:
\begin{itemize}
\item{} We proposed an integration model with combining the network embedding and the integrating learning, which can perform link prediction and sign prediction in the same framework.

\item{} In spite of the generality, the integration model can generate better performance than the special methods such as \textit{SEAL} \cite{zhang2018link} for link prediction and \textit{Beside} \cite{chen2018bridge} for sign prediction.

\item{} We found that only low-dimensional embedding can achieve very good performance in the integration model.
\end{itemize}

	The rest of this paper is organized as follows. Sec. \ref{sec:relatedwork} discusses related work. We describe our method in Sec. \ref{sec:proposed}. Sec. \ref{sec:experiment} provides information on our experiment and its results. In Sec. \ref{sec:conclusion}, we summarize the paper and discuss prospects for future work.
	
	\section{Related Work}
	
	\label{sec:relatedwork}
	Liben-Nowell and Kleinberg first proposed the concept of link prediction \cite{liben2007link}, i.e., to predict missing or hidden links through network topology and node information, which can be solved by the similarity of node pairs, where higher similarity implies a link is more likely. Zhang et al. experimentally analyzed the robustness  to random noise of several similarity metrics in link prediction \cite{zhang2016measuring}. Lu et al. proposed a method to predict drug-target interaction (DTI) based on  network topology, showing that the Jaccard index is more effective than the restricted Boltzmann machine (RBM) \cite{lu2017link}. Some current work is no longer based on the similarity index. Rawan et al. converted  link prediction to a classification problem, and they developed evolutionary neural network-based models to solve it \cite{yaghi2020link}. They found that results of the prediction algorithms based on similarity measures, such as common neighbors and the Adamic/Adar index, have low accuracy because they depend on the application domain. This paper treats link prediction as a classification problem, and solves it by supervised machine learning. Based on graph convolutional networks (GCNs) in modeling graph data, TransGCN \cite{cai2019transgcn} simultaneously learns relation and entity embeddings. DeepLinker deploys a graph attention network \cite{gu2019link}, and it uses links as supervising information for link prediction instead of learning node representation from node label information. Besides the link prediction algorithms, the perturbation of the adjacency matrix \cite{lu2015pnas} and the normalized shortest compression length\cite{sun2020nc} are adopted to study the link predictability of complex networks.
	
	Similar to link prediction, sign prediction can also be based on network similarity. Symeonidis et al. proposed similarity measures for nodes residing in the same or different communities \cite{symeonidis2013spectral}, and introduced metrics that consider status theory for computation of node similarities. Most sign prediction methods assume that signs in adjacent edges have certain similarities, and some models rely on more complex sign theories to infer the spread of edge information. Such methods require higher network sampling rates and may fail at low sampling levels \cite{naaman2019edge}. Similar to  methods based on random walk, they need not consider the complex substructure in a network. The embedded features of a network after the walk already contain its topological characteristics so as to accommodate complex network structures with different properties. This problem can also be addressed by supervised machine learning. DuBois et al. used more features to train the model for improved sign prediction \cite{dubois2011predicting}. Information such as gender, interest, and location of nodes in social networks have been used as training features. This enrichment of features enables more information to be captured by a model, which is an advantage of feature engineering. Graph autoencoders (AEs) and variational autoencoders (VAEs) have been utilized in link prediction \cite{salha2019keep}, and the gravity-inspired graph AE extends these to solve the sign prediction problem \cite{salha2019gravity}. Some models explore user personality as  complementary information from social media \cite{beigi2019signed}.
	
	The balance theory describes the characteristics of a triangular network structure, and it is widely used in  sign prediction \cite{patidar2012predicting}, where ``friends of friends are my friends, friends of enemies are my enemies, friends of enemies are my enemies, and enemies of enemies are my friends." With preserving the structural balance, Shen and Chung \cite{Shen2020tc} proposed a deep network embedding model, which employs a semisupervised stacked auto-encoder to conduct the link sign prediction. Chen et al. \cite{chen2018bridge} proposed the \textit{Beside} method, which performed  well on sign prediction, considering both triangular and bridge structures in a network. Although social balance and status theory have attracted much attention in sign prediction, Tang et al. noted that  network data are sparse, and the combination of triangular structural features may not be suitable \cite{tang2016survey}. Some methods are based on the Laplace spectrum method and network embedding. With the vigorous development of machine learning and deep learning, network embedding methods show better performance. We study the problem of link and sign prediction based on network embedding in this work.

	\section{Proposed Methods}
	\label{sec:proposed}
	\subsection{Problem Definition}
	
	We treat link and sign prediction as binary classifications across network data, where the network is defined as follows.
	
	\begin{itemize}
		\item{{\bf link prediction}}: An unsigned graph, as in Fig. \ref{fig:framework}(A), can be represented by ${G(V,E)}$, where ${\bm{v}_i\in V}$ and ${\bm{e}_{ij}\in E}$ denote the node and edge set, respectively. If ${\bm{e}_{i,j}=1}$, then there is an edge between nodes ${\bm{v}_i}$ and ${\bm{v}_j}$, and otherwise ${\bm{e}_{i,j}=0}$. The link prediction task is to find a potential edge based on the observed network structure, as in Fig. \ref{fig:framework}(F).
		
		\item{{\bf sign prediction}}: A signed directed graph, as in Fig. \ref{fig:framework}(B), can be represented by $G_s(V,E_s)$, where ${\bm{v}_i \in V}$ and ${\overrightarrow{e_{ij}} \in E_s}$ are the node and edge set, respectively. If $\overrightarrow{e_{ij}}=+/-1$, then  there is a positive (negative) tendency between nodes ${v_i}$ and ${v_j}$. The sign prediction task is to find the potential edge sign ($+/-1$) based on the observed information, as shown in Fig. \ref{fig:framework}(G).
	\end{itemize}
	
	\subsection{Model Description}
	
	\subsubsection{Model Framework}
	We propose an integration model for both link and sign prediction, whose overall architecture is shown in Fig. \ref{fig:framework}. Firstly, we represent the network nodes as a low-dimensional vector using the network embedding approaches, and then some feature engineering methods can be used to generate more useful features for each node pair. Then an integrated learning, which can obtain the classification results by combining many basic learners, is applied to obtain the link prediction. Therefore, the proposed method has three main components. (i) \textbf{Network Embedding} (Fig. \ref{fig:framework}(A)-C)): for each node in the network, we need to obtain a low dimensional vector in the vector space by using the representation learning algorithm on the network, which is also referred to as network embedding. (ii) \textbf{Feature Engineering} (Fig. \ref{fig:framework}(D)): in order to predict the link or the sign between the node pairs well, we need to obtain the better training features based on the node vector using the feature engineering, which can also improve the upper score limit of the prediction model. (iii) \textbf{Integrated Learning} (Fig. \ref{fig:framework}(E)-(G)): integrated learning is used to process high-dimensional complex features, and the obtained classifiers are used for both link and sign prediction.
	
	\subsubsection{Network Embedding}
	
	As shown in Fig. \ref{fig:framework}(C), the network embedding model can learn network data topology features, and each node  is represented by a ${k}$-dimensional vector. This method can capture the network structure features and transform network data to a matrix of  fixed shape. Especially, for the sign prediction of the signed network, we ignored the sign to get the node representation. And the sign information was used as the label of the node pairs in the classification training. The integrated model supplements the missing sign features of the network so that it can learn the sign features and make up for the shortcomings of the network embedding. In this way, the generalization ability of the integrated model to the network data is greatly improved.
	
	We take  Node2Vec \cite{grover2016node2vec}, which is based on a biased random walk, as an example to learn the node features. Given a node ${v}$, the probability of choosing the next node ${x}$   is
	\begin{equation}
	\label{equ:equ3}
	P(c_i=x\vert c_{i-1}=v)=\left\{\begin{array}{l}\frac{\pi_{vx}}Z\;\;if\;\;(v,x)\in E\\0\;\;\;\;otherwise\end{array}\right.,
	\end{equation}
	where ${\pi_{vx}}$ is the transition probability between nodes ${v}$ and ${x}$, and ${Z}$ is a normalization constant. Suppose the current random walk passes the edge ${(t, v)}$ to reach  node ${v}$. Let ${\pi_{vx}=\alpha_{pq}(t,x)\cdot w_{vx}}$, where ${\alpha_{pq}(t,x)}$ is the probability that node ${x}$ deviates from node ${t}$, $w_{vx}$ is the edge weight between nodes ${v}$ and   ${x}$, and node ${t}$ comes before ${v}$ in the random walk sequence. Then ${\alpha_{pq}(t,x)}$ can be obtained as
	
	\begin{equation}
	\label{equ:equ4}
	\alpha_{pq}\left(t,x\right)=\left\{\begin{array}{l}\frac1p\;\;if\;d_{tx}=0\\1\;\;\;if\;d_{tx}=1\\\frac1q\;\;if\;d_{tx}=2\end{array}\right.,
	\end{equation}
	where ${d_{tx}}$ is the shortest path distance between nodes ${t}$ and ${x}$. ${\alpha=1}$ when the distance between the next node ${x}$ and the previous node ${t}$  equals the distance between node ${x}$ and the current node ${v}$; ${\alpha=1/p}$ when the next node ${x}$ is the previous node; and otherwise ${\alpha=1/q}$. The parameters $p$ and $q$ control the jump probability of the random walk sequence. After obtaining a series of random walk sequences, Node2Vec uses the Skip-Gram model to train these node sequences, and generates a corresponding vector for each node in the network.

	\begin{figure}[htbp]
		\centerline{\includegraphics[width=9cm]{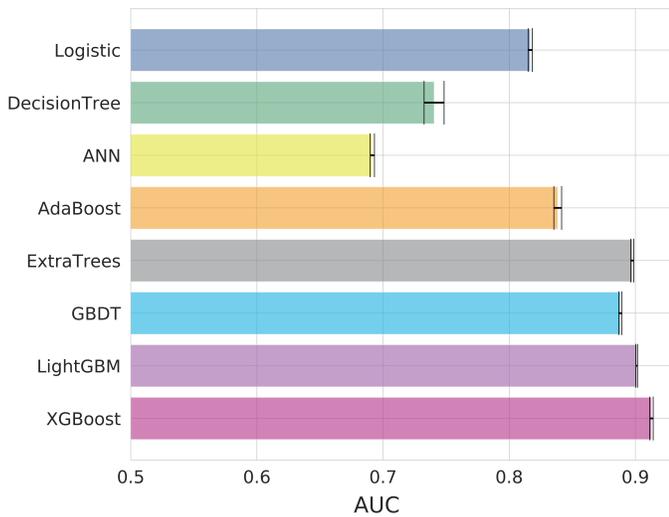}}
		\caption{Results of multiple basic classifiers in K-fold cross-validation, which are used to select  higher scores and less volatile classifiers. AUC is used to evaluate the effectiveness of each classifier. The length of the histogram represents the average value of K-fold cross-validation scores. The error bar of each classifier in K evaluations can also be obtained. The results are performed on the Slashdot dataset.}
		\label{fig:k-flod}
	\end{figure}
	
	\begin{figure}[htbp]
		\centerline{\includegraphics[width=9cm]{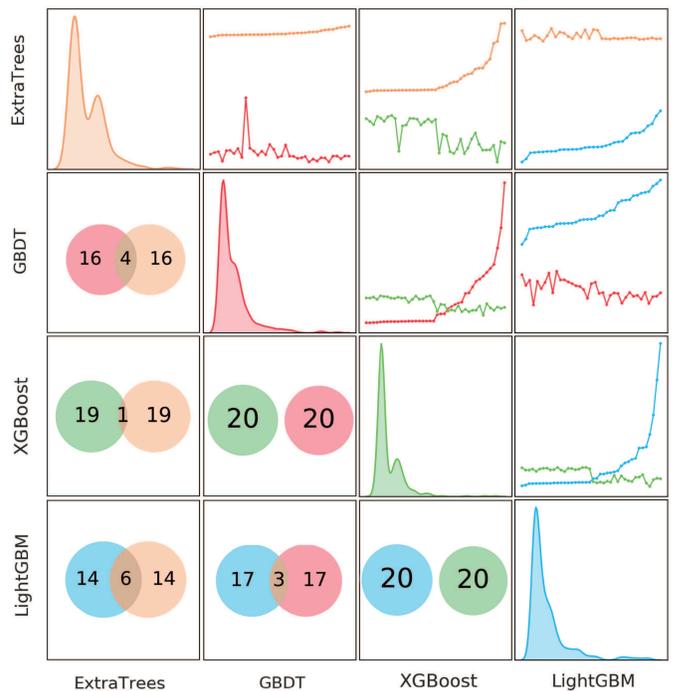}}
		\caption{The analysis of the important features of the four basic classifiers. Pink, red, green and blue represent ExtraTrees, GBDT, XGBoost and LightGBM respectively. The diagonal subgraphs illustrate the feature importance distribution of the four models, where the x-axis represents the value of importance, and the y-axis represents the number of features with the corresponding importance. Sub-graphs in the lower-left part show Venn diagrams of the top 20 most important features of paired classifiers, where the overlap between two classifiers of the top 20 features is small. Sub-graphs in the upper-right part show the correlation of the feature importance between the paired classifiers, where x-axis represents the index of the features of the corresponding Venn diagram, and y-axis represents the corresponding feature importance for different basic classifier. These subplots show that each basic classifier successfully learned different features. The results are performed on the Slashdot dataset.}
		\label{fig:feature}

	\end{figure}
	
	\subsubsection{Feature Engineering}
	
	Nodes that are close in a network tend to be near each other in embedding space, and the distance between nodes in the feature space, such as nodes $V_s$ and $V_e$ in Fig. \ref{fig:framework}(D), is a good indicator of their relationship or relative status, which should be an effective feature in link and sign prediction. After extensive experiments and analyses, we deploy a classical feature strengthening technique that concatenates the features of two nodes and their distance features ($\overrightarrow{V_s}$, $\overrightarrow{V_e}$ and $\overrightarrow{V_s}-\overrightarrow{V_e}$ in Fig. \ref{fig:framework}(D)). In the experiment, we set the node embedding dimension to 128, and the feature dimension after feature engineering was ${128*3}$. Using this feature representation, the training and validation sets can be constructed. In this feature style, the attribution information of nodes is preserved as supplementary knowledge to support the prediction. The experimental results show that the extended features effectively improve the model's accuracy.
	
	\subsubsection{Integrated Learning}
	
	After raw feature concatenation, we applied the integrated learning model to generate the link or sign predictions. In the integrated learning model, we firstly used several basic classifiers to train the raw features of the node pairs, and then used a meta-classifier to train the inferred results that obtained by several basic classifiers to obtain the final prediction. The meta-classifier generally performs well on both small- and large-scale networks (over 100,000-level edges or nodes), and has small score fluctuations for different feature dimensions. The chosen classifiers are important in the integration model. In the integration training, we need to analyze the performance differences between the basic classifiers and select  several basic classifiers with the highest score as the dominant input of the meta-classifier, since the basic classifiers with less differences will not improve the model performance or even introduce unnecessary redundancy. In this work, we use ExtraTree  \cite{geurts2006extremely}, GBDT  \cite{friedman2001greedy}, LightGBM \cite{ke2017lightgbm}, and XGBoost \cite{chen2016xgboost} as the basic classifiers (see Fig. \ref{fig:framework}(E)).
	
	Taking the Slashdot \cite{leskovec2010signed} dataset as an example, we partitioned the training and validation sets into $K$ subsets and used  cross-validation  to select a basic classifier, as shown in Fig. \ref{fig:framework}(E). We applied five-fold cross-training on several classifiers on the training set, with results as shown in Fig. \ref{fig:k-flod}. The classifiers performed quite differently, and   ExtraTree, GBDT, LightGBM, and XGBoost were much better than the others (AUC values  greater than 0.85). The error bars of these four basic classifiers are relatively low, hence they should generate more stable prediction in the final integration model. The high AUC values of the basic-models showed that the four basic-models have extracted the efficient features. Therefore, we don’t need the complex model to integrate the basic-models. In comparison, we chose the Random Forest as the meta-classifier,  because it is easier to avoid over-fitting and doesn’t need feature dimension reduction. We also checked the performance differences between the four basic classifiers.  Fig. \ref{fig:feature} shows the correlation between the 20 most important features of each classifier across various classifiers. Generally, the feature importance of the tree-based models can be quantified by the number of times that the corresponding feature is used to split the data across all trees. There was little overlap of different classifiers (Venn diagram in Fig. \ref{fig:feature}), and the importance of these features was  very different (dotted-line graph in Fig. \ref{fig:feature}). The differences between the most important features among the classifiers indicated that the contribution of the features to the prediction of four classifiers was very different. The  integration of these classifiers would combine these contributions from various aspects to improve the prediction performance.

	\subsection{Model Training}
	
	Basic classifiers must cooperate in the integrated model for better performance, and selecting proper hyperparameters is critical to performance. From the tree model to the neural network model, many hyperparameters are significantly correlated with the model effect, such as network depth, learning rate, tree depth, number of leaf nodes, and regularization. To find the best combination of hyperparameters requires not only  model construction experience but hyperparameter optimizers for auxiliary parameter adjustment. We use K-fold cross-validation and some automatic parameter adjustment methods during the training process. Training checks the model for overfitting, underfitting, just-fitting, training shock, and complete non-convergence. The performance of the last two trainings rarely appears in the experiment without detailed explanation.
	
	\begin{itemize}
		\item{{\bf Just-fitting}}: The model may have achieved good results. It is necessary to reduce the number of leaf nodes and network layers  to reduce the model complexity, amount of calculation, and required calculation space.
		
		\item{{\bf Overfitting}}: The model has converged on the training set, but the performance of the validation set is not good. These are   manifestations of  poor generalization ability. We use many methods to avoid overfitting, including reducing the height of the tree model,  number of leaf nodes, and learning rate, along with L1 and L2 regularization  \cite{micchelli2005learning}.
		
		\item{{\bf Underfitting}}: The model has no obvious convergence effect on the training and validation sets. The problem can be solved through conventional methods such as to increase the number of iterations and the model complexity, and to optimize the dataset. We have tried distance-based fake data construction methods such as ADASYN \cite{he2008adasyn}, which are more reasonable to increase the proportion of positive and negative samples.
	\end{itemize}
	
	In model training, we fix some important hyperparameters by limiting the parameter range through the performance of K-fold cross, and fine-tune them through hyperparameter optimization methods including grid search based on exhaustive search. We mainly adopt Bayesian optimization, with parameter dimension of about 20. In  Bayesian parameter tuning, we adopt a Gaussian process, with considering the previous parameter information, constantly updating the prior, fewer iterations and not easy to fall into local optimal conditions. Take the Slashdot dataset as an example, the settings of main hyperparameters are shown in Table \ref{tab:baseline_parameter} .
	
	\begin{table*}[htb]
		\centering
		\caption{Settings of main hyperparameters for the Slashdot dataset, where "-" indicates that the method has no corresponding parameters or the model does not use the relevant parameters.}
		\begin{tabular}{lllllll}
			\toprule
			parameter  & AdaBoost & DecisionTree & ExtraTrees & GBDT & LightGBM & XGBoost \\
			\midrule
		    max\_depth & 8 & 10 & 15 & 7 & 7 & 16 \\
			min\_samples\_leaf & 7 & 5 & 3 & 5 & - & - \\
			min\_samples\_split & 4 & 3 & 3 & 2 & - & - \\
			colsample\_bytree & - & - & - & - & 0.47 & 0.85 \\
			n\_estimators & 30 & - & 300 & - & 800 & 512 \\
			learning\_rate & 0.05 & - & - & 0.1 & 0.3 & 0.01 \\
			reg\_alpha & - & - & - & - & 0.5 & 0.3 \\
			reg\_lamdba & - & - & - & - & 0.1 & 0.243 \\
			subsample & - & - & - & - & 0.567 & 0.85 \\
			\bottomrule
		\end{tabular}%
		\label{tab:baseline_parameter}%
	\end{table*}%

	\begin{figure}[htb]
		\centerline{\includegraphics[width=9cm]{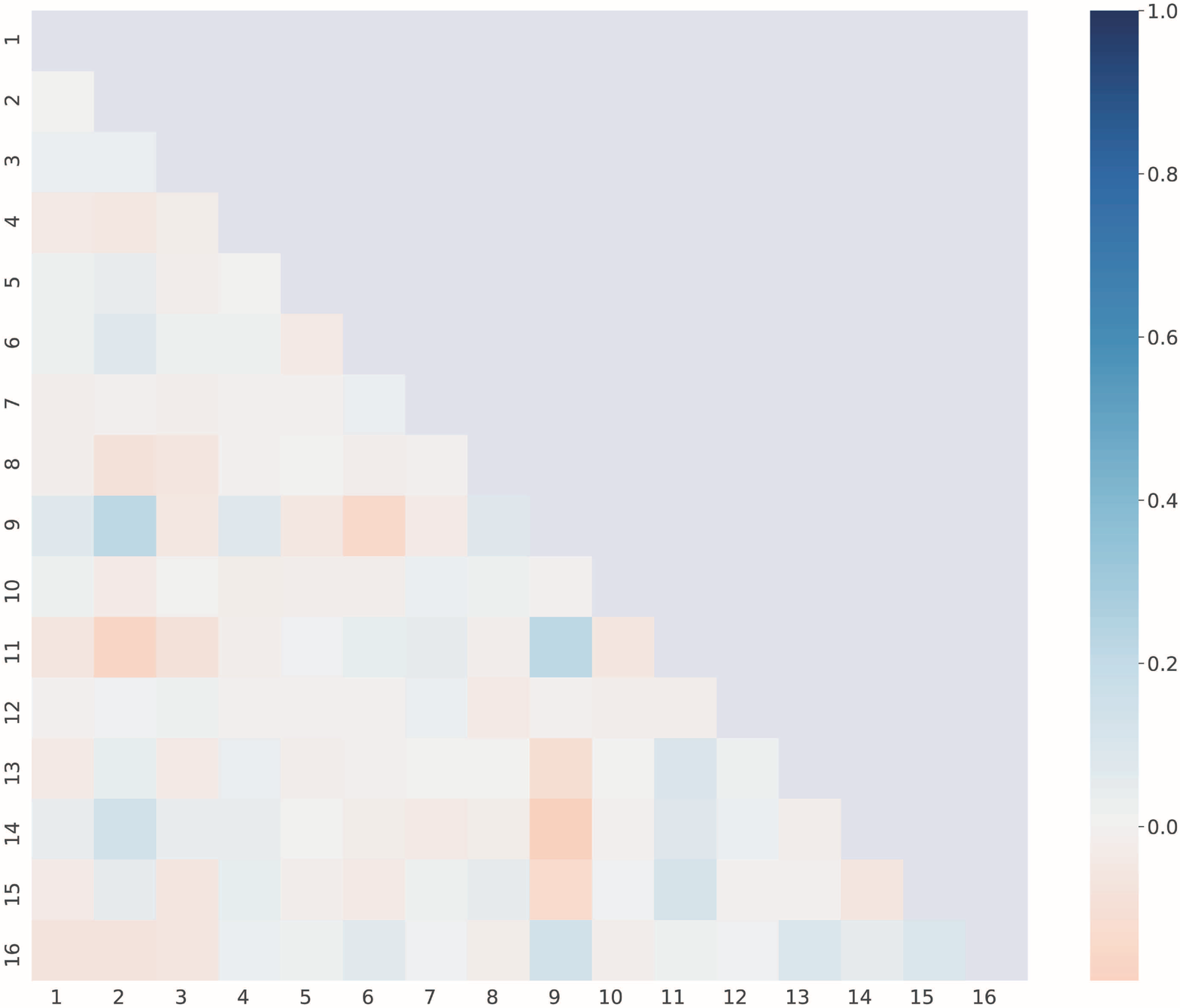}}
		\caption{The heat map shows the Pearson correlation between two-dimensional features. For convenience, we set the node representation dimension of the network embedding model to 16. For example, 1 on the abscissa and 16 on the ordinate indicate that the correlation coefficient between the vector composed of the first dimension of all node embedding vectors and the vector composed of the 16th dimension is -0.07. A correlation coefficient closer to 1 indicates a stronger  two-dimensional positive correlation, and the color of the table is darker. A correlation coefficient closer to zero indicates less correlation between dimensions, and the color is lighter. In the same way, if there is a negative correlation, the correlation coefficient is closer to -1. Almost no correlation is seen between various dimensions, so there is no status of data redundancy. This occurs regardless of the network embedding dimension, which indicates that the node feature data after network embedding can be directly used to construct the sample set, with no post-processing step of PCA dimensionality reduction. The results are performed on the Slashdot dataset.}
		\label{fig:heatmap}
	\end{figure}

	\section{Experiment}
	\label{sec:experiment}
	
	\subsection{Data description}

	\begin{table}[htbp]
		\centering
		\caption{Basic statistics of datasets.  \#$\bar{k}$ represents the number of the average degree. \#+$(\%)$ and \#-$(\%)$ represent the percentages (\%) of the edges labeled + and - in the signed network respectively.}
		\begin{tabular}{p{4.19em}ccccc}
			\hline
			Dataset & \#Nodes & \#Edges & \#$\bar{k}$ & \#+(\%) & \#-(\%)\\
			\hline
			Celegans & 297   & 2,148 & 14.46  & & \\
			PB    & 1,222 & 16,714 & 27.36 & & \\
			Yeast & 2,375 & 11,693 & 9.85 & & \\
			Facebook & 4,039 & 88,234 & 43.69 & & \\
			Slashdot & 82,140 & 549,202 & 13.37 & 77.4  & 22.6 \\
			Epinions & 131,828 & 841,372 & 12.76 & 85.3  & 14.7 \\
			Wikirfa & 11,258 & 179,418 & 31.87 & 77.92 & 22.08 \\
			\hline
		\end{tabular}%
		\label{tab:data}%
	\end{table}%
	
	We selected seven real-world graph datasets on which to conduct experiments to check the performance of the  integration model. Four datasets were used for link prediction, and three datasets for sign prediction. (i) \textbf{Celegans} \cite{watts1998collective} is a neural network of c. elegans. (ii) \textbf{PB} \cite{ackland2005mapping} is a network of U.S. political blogs. (iii) \textbf{Yeast} \cite{von2002comparative} is a protein-protein interaction network in yeast. (iv) \textbf{Facebook} \cite{leskovec2016snap} is a social network extracted from Facebook. (v) \textbf{Slashdot} \cite{leskovec2010signed} is a technology-related news site where users can mark each other as friend or foe. (vi) \textbf{Epinions} \cite{leskovec2010signed} is an online social network and consumer review site whose members can decide whether to ``trust" each other. (vii) \textbf{Wikirfa} \cite{west2014exploiting} records the voting process during ``request for adminship (RfA)," where any community member can cast a supporting, neutral, or opposing vote for a Wikipedia edit. The basic statistics are summarized in Table~\ref{tab:data}.
	
	We consider both link and sign prediction as binary classification problems. For link prediction, the positive samples are the observed edges, and the negative samples are   unconnected node pairs randomly sampled from the training sets. For sign prediction, positive samples are edges labeled with "+" (e.g., friend, trust), and negative samples are labeled with "-" (e.g., foe, distrust). We extracted similar numbers of positive and negative samples in model training.
	
\subsection{Feature Analysis}
	
\begin{figure*}[htbp]
\centerline{\includegraphics[height=12cm,width=13cm]{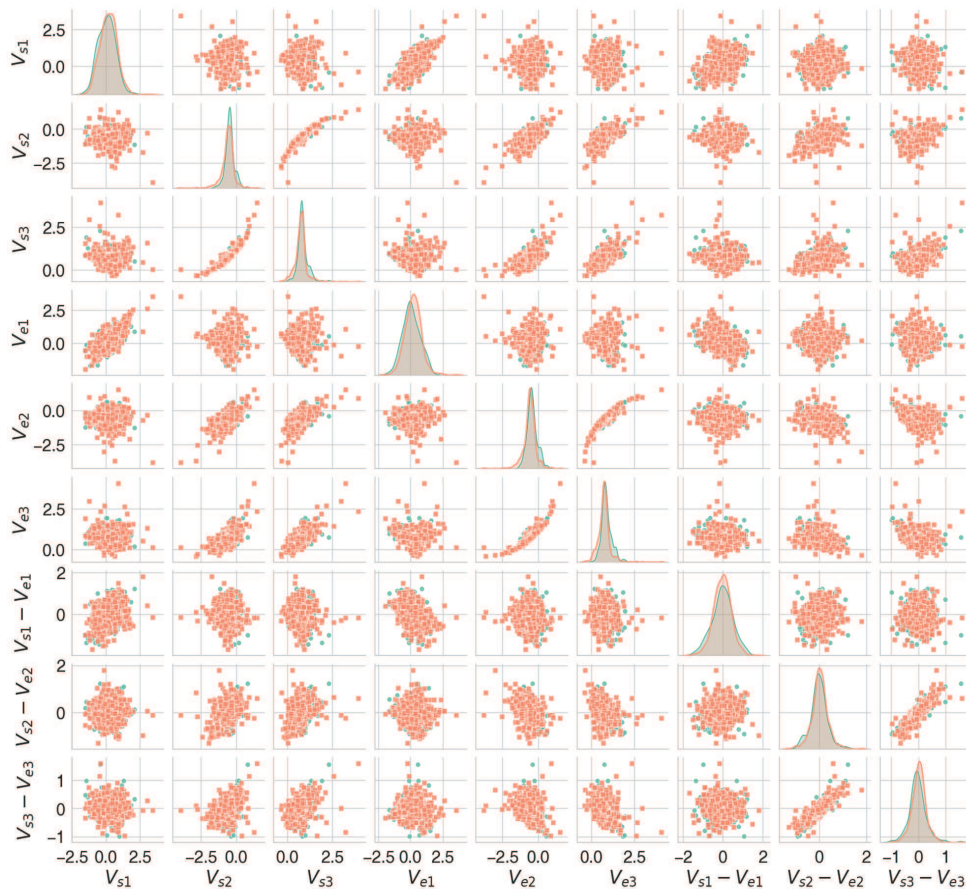}}
\caption{The pair plot provides  reasonable  insight into distribution features. The case of a three-dimensional vector is illustrated. Colors correspond to labels in the training set; orange and aquamarine indicate negative and positive samples, respectively. From the diagonal subgraphs, we can see that these features are close to normally distributed, which is consistent with the assumption of linear regression. From the scatter plot, outliers are small compared to the total amount of training data, and the red and blue points are heavily coincident on the subgraph. Simple classifiers cannot accurately distinguish training data for some perspectives. Another interesting finding is between $ V_{s3} $ and $ V_{s2} $, and between $ V_{e3} $ and $ V_{e2} $. From  the point cloud of the subgraph, one almost sees  an exponential function, although this cannot explain the   meaning of the embedded node features, which shows that Node2vec has some interpretability. The results are performed on the Slashdot dataset.}
\label{fig:pairplot}
\end{figure*}
	
We analyzed the node feature data obtained by the network embedding model, and measured the correlation between the two embedding dimensions of the network through the Pearson correlation coefficient, thereby judging the quality of the feature obtained by the network embedding model. To facilitate the presentation, we set the node embedding dimension to 16. In Fig. \ref{fig:heatmap}, we can see that the two most relevant sets of dimensions are the second and ninth dimensions and   ninth and eleventh dimensions, respectively, and the correlation coefficient dimension is only 0.2. The correlation between most dimensions only slightly fluctuates around 0, which indicates that the correlation between the dimension features obtained by the network embedding model is very low, and the independence of dimensions is strong. As the network embedding dimension increases or decreases, this phenomenon still exists, which indicates that the node feature data after network embedding do not require the processing of PCA and other dimensionality reduction operations.
	\begin{table*}[htb]
		\centering
		\caption{Results of Link Prediction. ``*" Indicates   Input Features of the Model Are Obtained From Network Feature Engineering. Bolded Scores   Indicate the Best Performance of All Methods, and Underlined Scores   Denote the Best Performance of All Baselines.}
		\begin{tabular}{lp{4.19em}llllllllll}
			\toprule
			& Metric & \multicolumn{1}{p{2.19em}}{DW} & \multicolumn{1}{p{3.19em}}{LINE} & \multicolumn{1}{p{5.315em}}{Graph\newline{}Factorization} &
			\multicolumn{1}{p{3.19em}}{SEAL} & \multicolumn{1}{p{3.69em}}{Logistic} & \multicolumn{1}{p{4.19em}}{Logistic*} & \multicolumn{1}{p{3.19em}}{ANN} & \multicolumn{1}{p{3.19em}}{ANN*} &
			\multicolumn{1}{p{4.19em}}{Integration Model} & \multicolumn{1}{p{4.19em}}{Integration Model*} \\
			\midrule
			\multicolumn{1}{c}{\multirow{4}[1]{*}{Celegans}} & AUC  & 0.7992 & 0.7215 & 0.7278 & 0.9036 & 0.8132 & 0.8599 & 0.8870 & 0.8950 & \underline{0.9515} & \textbf{0.9735} \\[0.6em]
			& macro-F1 & 0.7205 & 0.6576 & 0.6623 & 0.8033 & 0.7343 & 0.7718 & 0.8473 & \underline{0.8950} & 0.8847 & \textbf{0.9115} \\[0.6em]
			& micro-F1 & 0.7206 & 0.6577 & 0.6624 & 0.8047 & 0.7346 & 0.7718 & 0.8475 & \underline{0.8952} & 0.8847 & \textbf{0.9115} \\[0.6em]
			& binary-F1 & 0.7206 & 0.6533 & 0.6564 & 0.7868 & 0.7462 & 0.7710 & 0.8518 & \underline{0.8996} & 0.8830 & \textbf{0.9130} \\[0.6em]
			\midrule
			\multicolumn{1}{c}{\multirow{4}[0]{*}{PB}} & AUC   & 0.8404 & 0.7942 & 0.7247 & 0.9370 & 0.8456 & 0.9490 & 0.9208 & 0.9583 & \underline{0.9840} & \textbf{0.9911} \\[0.6em]
			& macro-F1 & 0.7573 & 0.7256 & 0.6722 & 0.8666 & 0.7604 & 0.8837 & 0.9409 & \textbf{0.9582} & 0.9195 & \underline{0.9539} \\[0.6em]
			& micro-F1 & 0.7574 & 0.7257 & 0.6723 & 0.8668 & 0.7604 & 0.8837 & 0.9409 & \textbf{0.9582} & 0.9195 & \underline{0.9539} \\[0.6em]
			& binary-F1 & 0.7593 & 0.7297 & 0.6771 & 0.8706 & 0.7617 & 0.8842 & 0.9196 & \textbf{0.9585} & 0.9409 & \underline{0.9546} \\[0.6em]
			\midrule
			\multicolumn{1}{c}{\multirow{4}[0]{*}{Yeast}} & AUC   & 0.6836 & 0.6935 & 0.6407 & 0.9659 & 0.6717 & 0.8413 & 0.9625 & 0.9803 & \underline{0.9970} & \textbf{0.9974} \\[0.6em]
			& macro-F1 & 0.6907 & 0.6483 & 0.6072 & 0.9153 & 0.6461 & 0.7710 &0.9624 & 0.9803 & \underline{0.9850} & \textbf{0.9867} \\[0.6em]
			& micro-F1 & 0.6928 & 0.6485 & 0.6077 & 0.9153 & 0.6474 & 0.7710 & 0.9624 & 0.9803 & \underline{0.9850} & \textbf{0.9867} \\[0.6em]
			& binary-F1 & 0.7157 & 0.6574 & 0.6206 & 0.9157 & 0.6675 & 0.7722 & 0.9623 & 0.9807 & \underline{0.9853} & \textbf{0.9870} \\[0.6em]
			\midrule
			\multicolumn{1}{c}{\multirow{4}[1]{*}{Facebook}} & AUC   & 0.9200  & 0.8100  & 0.8288 & \underline{0.9936} & 0.8874 & 0.9169 & 0.9690 & 0.9769 & 0.9935 & \textbf{0.9954} \\[0.6em]
			& macro-F1 & 0.9083 & 0.8457 & 0.7738 & 0.9650 & 0.8686 & 0.8611 & 0.9690 & \underline{0.9769} & 0.9755 & \textbf{0.9800} \\[0.6em]
			& micro-F1 & 0.9088 & 0.8465 & 0.7745 & 0.9650 & 0.8694 & 0.8614 & 0.9690  & \underline{0.9769} & 0.9755 & \textbf{0.9800} \\[0.6em]
			& binary-F1 & 0.9150 & 0.8573 & 0.7860 & 0.9651 & 0.8789 & 0.8681  & 0.9691 & \underline{0.9770} & 0.9758 & \textbf{0.9803} \\[0.6em]
			\bottomrule
		\end{tabular}%
		\label{tab:result1}%
	\end{table*}%
	\begin{table*}[htb]
		\centering
		\caption{Results for Sign Prediction. ``*" Indicates That Network Feature Engineering is Deployed. Scores in Bold Indicate the Best Performance of All Methods, and Scores with Underlines Denote the Best Performance of All Baselines.}
		\begin{tabular}{cp{4em}ccccccccccc}
			\toprule
			& Metric & \multicolumn{1}{p{2.75em}}{DW} & \multicolumn{1}{p{2.5em}}{LINE} & \multicolumn{1}{p{2.565em}}{SIDE} &
			\multicolumn{1}{p{4em}}{Beside-tri} & \multicolumn{1}{p{2.69em}}{Beside} & \multicolumn{1}{p{2.875em}}{Logistic} &
			\multicolumn{1}{p{4em}}{Logistic*} & \multicolumn{1}{p{2.75em}}{ANN}& \multicolumn{1}{p{2.75em}}{ANN*} & \multicolumn{1}{p{3.5em}}{Integration Model} &
			\multicolumn{1}{p{4em}}{Integration Model*} \\
			\midrule
			\multicolumn{1}{c}{\multirow{4}[1]{*}{Slashdot}} & AUC   & 0.7743 & 0.5589 & 0.8495 & 0.8759 & \underline{0.9092} & 0.5735 & 0.8172 & 0.6742 & 0.7260 & 0.7300  & \textbf{0.9122} \\[0.6em]
			& macro-F1 & 0.5994 & 0.4368 & 0.7433 & 0.7594 & \underline{0.7985} & 0.4364 & 0.6585 & 0.4763 & 0.7501 & 0.5608 & \textbf{0.8013} \\[0.6em]
			& micro-F1 & 0.7779 & 0.7740 & 0.8407 & 0.8457 & \underline{0.8649} & 0.7743 & 0.8029 & 0.8135 & 0.8443 & 0.7891 & \textbf{0.8701} \\[0.6em]
			& binary-F1 & 0.8668 & 0.8726 & 0.9015 & 0.9035 & \underline{0.9142} & 0.8728 & 0.8806 & 0.8715 & 0.9035 & 0.878 & \textbf{0.9182} \\[0.6em]
			\midrule
			\multicolumn{1}{c}{\multirow{4}[0]{*}{Epinions}} & AUC   & 0.8170 & 0.6012 & 0.8730 & 0.9304 & \underline{0.9439} & 0.6959 & 0.8574  & 0.7460 & 0.7971 & 0.8200  & \textbf{0.9520} \\[0.6em]
			& macro-F1 & 0.6141 & 0.8543 & 0.8223 & 0.8478 & \underline{0.8679} & 0.4796 & 0.7329 & 0.7931 & 0.8374 & 0.7008 & \textbf{0.8741} \\[0.6em]
			& micro-F1 & 0.8696 & 0.9213 & 0.9234 & 0.9306 & \underline{0.9376} & 0.8523 & 0.8923 & 0.8718 & 0.9283 & 0.8835 & \textbf{0.9421} \\[0.6em]
			& binary-F1 & 0.9279 & 0.9213 & 0.9564 & 0.9600  & \underline{0.9638} & 0.9200  & 0.9392 & 0.9298 & 0.9589 & 0.9346 & \textbf{0.9667} \\[0.6em]
			\midrule
			\multicolumn{1}{c}{\multirow{4}[1]{*}{Wikirfa}} & AUC  & 0.6672 & 0.5808 & 0.7715 & 0.8668 & \underline{0.8814} & 0.6281 & 0.7785 &0.6073 & 0.7213 & 0.7733 & \textbf{0.9027} \\[0.6em]
			& macro-F1 & 0.6204 & 0.5574 & 0.7170 & \underline{0.7844} & \textbf{0.8013} & 0.4530 & 0.6367 & 0.6248 & 0.7433 & 0.6372 & 0.7790 \\[0.6em]
			& micro-F1 & 0.6050 & 0.5575 & 0.7172 & 0.7845 & 0.8013 & 0.7826 & 0.8097 & 0.8057 & \underline{0.8425} & 0.8103 & \textbf{0.8611} \\[0.6em]
			& binary-F1 & 0.8773 & 0.8761 & 0.8917 & 0.9084 & \underline{0.9097} & 0.8777 & 0.8874 & 0.8853 & 0.9029 & 0.8878 & \textbf{0.9137} \\[0.6em]
			\bottomrule
		\end{tabular}%
		\label{tab:result2}%
	\end{table*}%
	
To  analyze the distribution of high-dimensional network embedding features, we randomly extracted 1000 samples with an embedding dimension of 3 from the Slashdot training dataset, and we show these features using a pair plot. In Fig. \ref{fig:pairplot}, $ V_{s1} $, $ V_{s2} $, $ V_{s3} $ represent a three-dimensional feature of the starting node, $ V_{e1} $, $ V_{e2} $, $ V_{e3} $ represent a three-dimensional feature of the ending node, and $ V_{s1}-V_{e1} $, $ V_{s2}-V_{e2} $, $ V_{s3}-V_{e3} $ represent a three-dimensional feature obtained by the starting node vector minus the ending node vector. We find that the target variable is normally distributed, which validates the linearity of the models. We can also find that there are few outliers in the point graph through these two features, indicating that the network embedding method has performed well to some extent, and no more preprocessing is required.
	
\subsection{Comparison Methods}
	
	We compared the prediction performance of the  integration model on seven datasets to that of the following  methods.
	
	\begin{itemize}
		\item{} DeepWalk (\textit{DW}) \cite{perozzi2014deepwalk} is network embedding method, which is based on the Skip-gram language model. It cannot distinguish positive and negative edges, so the sign information was discarded during the random walk in sign prediction.
		
		\item{} LINE \cite{tang2015line} is a network embedding method, which considers first- and second-order proximity. It  cannot distinguish between positive and negative edges, and the preprocessing of sign prediction is similar to that of DeepWalk.
		
		\item{} SIDE \cite{kim2018side} is a signed network embedding method based on random walks. It aggregates signs and directions along a path according to balance theory.
		
		\item{} Graph factorization \cite{ahmed2013distributed} is a framework for large-scale graph decomposition and inference.
		
		\item{} SEAL \cite{zhang2018link}  learns from local enclosing subgraphs, embeddings, and attributes based on graph neural networks.
		
		\item{} Beside\_tri \cite{chen2018bridge} incorporates the balance and status social-psychology theories to model triangle edges in a complementary manner. The triangle edge set $E_{bridge}$ consists of  edges whose adjacent nodes share at least one common neighbor.
		
		\item{} Beside \cite{chen2018bridge} incorporates the balance and status theories to model both triangle and bridge edges in a complementary manner. The bridge edge set $E_{bridge}$ consists of edges whose adjacent nodes share no common neighbors,
		\begin{equation}
		\label{equ:equ1}
		E_{triangle}=\{ \overrightarrow{e_{ij}} \vert N(v_i) \cap N(v_j)\neq \varnothing \}
		\end{equation}
		\begin{equation}
		\label{equ:equ2}
		E_{bridge}=\{ \overrightarrow{e_{ij}} \vert N(v_i) \cap N(v_j)= \varnothing \}.
		\end{equation}
		
		\item{} Like the integration model,  Logistic and ANN can also be adopted to train the node features generated by Node2vec. ANN includes three hidden layers and uses a dropout function to solve the problem of overfitting. Hidden layer parameters have uniformly distributed initialization. The number of neurons in each layer is 128, 32, 4, the dropout rate is 0.3, Adam is used as an optimizer, and the learning rate is 0.01.
		
	\end{itemize}
	
\subsection{Experimental Results}
	
\subsubsection{Experiments and Evaluation Metrics}
	
We evaluated  prediction ability by five-fold cross-validation. Datasets were randomly divided into five subsets, where four used as the training set and one for validation. The following metrics were used to evaluate link and sign prediction results \cite{islam2017distributed}\cite{leskovec2010predicting}\cite{wang2017signed}.
	
	\begin{itemize}
		\item{AUC} is the area under the receiver operating characteristic (ROC) curve. A global prediction performance indicator, it ignores sample imbalance and is expressed as
		\begin{eqnarray}
		\begin{split}
		\label{equ:equ5}
		AUC=\frac{\sum_{}^{}I(P_{pos}>P_{neg})}{M\ast N}
		\end{split},
		\end{eqnarray}
		where $\sum_{}^{}I(P_{pos}>P_{neg})$ is the number of sample pairs (positive  and  negative sample) in which the predicted probability of a positive sample is greater than that of a negative sample, and $M$ and $N$ respectively represent the amount of data for positive and negative samples. A  larger   value indicates better prediction accuracy.
		
		\item{Binary-F1}  comprehensively considers precision ($p$) and recall ($r$), and is defined as
		\begin{equation}
		\label{equ:equ6}	
		Binary\_F1=\frac{2p\times r}{p+r},
		\end{equation}
		where $p=\frac{TP}{TP+FP}$, $r=\frac{TP}{TP+FN}$ and $TP$, $FP$, and $FN$ are the numbers of true positive, false positive, and false negative samples respectively. A larger value indicates better prediction accuracy.
		
		\item{Macro-F1}  calculates the precision, recall, and F1-score of each category and obtains the F1-score of the entire sample by averaging. It does not consider the sample size of each category. It is calculated as
		\begin{eqnarray}
		\label{equ:equ7}	
		Macro\_F1=2\frac{Macro\_p\times Macro\_r}{Macro\_p+Macro\_r},
		\end{eqnarray}
		where $Macro\_p=\frac{\sum_i^n p_i}n$, $Macro\_r=\frac{\sum_i^n r_i}n$, $p_i$ and $r_i$ represent the precision and recall of category $i$ respectively, and $n$ is the number of label categories. A larger  value indicates better prediction accuracy.
		
		\item{Micro-F1}  does not need to distinguish categories, and it directly uses the precision and recall to calculate the F1-score. Mirco-F1 is more accurate for unbalanced category distribution, and is calculated as
		\begin{eqnarray}
		\label{equ:equ8}	
		Micro\_F1=2\frac{Micro\_p\times Micro\_r}{Micro\_p+Micro\_r},
		\end{eqnarray}	
		where $Micro\_p=\frac{\sum_i^nTP_i}{\sum_i^nTP_i+\sum_i^nFP_i}$ and $Micro\_r=\frac{\sum_i^nTP_i}{\sum_i^nTP_i+\sum_i^nFN_i}$. A  larger Micro-F1 value indicates better prediction accuracy.
	\end{itemize}

\subsubsection{Link Prediction}
	
Table \ref{tab:result1} compares the link prediction results of all approaches. We set the node embedding dimension to 128 for the logistic, ANN, and integration models. When marked with an asterisk in Table \ref{tab:result1} the input features of these models are generated by concatenating the feature vectors of the two nodes and the distances of the corresponding nodes ($\overrightarrow{V_s}$, $\overrightarrow{V_e}$ and $\overrightarrow{V_s}-\overrightarrow{V_e}$, the feature dimension is $128*3$). The largest AUC value (bolded in Table \ref{tab:result1}) for the integration model on all datasets indicates that it performed better  than the compared methods. For F1-related metrics, the integration model performed best on three datasets and competitively on the PB dataset, which is slightly weaker than ANN. The  results indicate that the integration model is a powerful methodology in link prediction.
	
The Logistic, ANN, and integration without ``*" indicate that the input features are just the distance of the corresponding nodes ($\overrightarrow{V_s}-\overrightarrow{V_e}$). Feature engineering can enhance link prediction, but the improvement is not so significant. For example, AUC increased from 0.9515 to 0.9735 for the Celegans data based on the integration model. This indicates that the feature of the distance between two nodes is very effective to infer the link between the corresponding nodes. The integration model without feature engineering performed even better than \textit{SEAL}, which is a neural network-based model that does  well at link prediction  \cite{zhang2018link}. (As illustrated in Table \ref{tab:result1}, \textit{SEAL} performed much better than  \textit{DW}, \textit{LINE}, and graph factorization in this experiment.)

\begin{figure*}[!htb]
\centerline{\includegraphics[height=9cm]{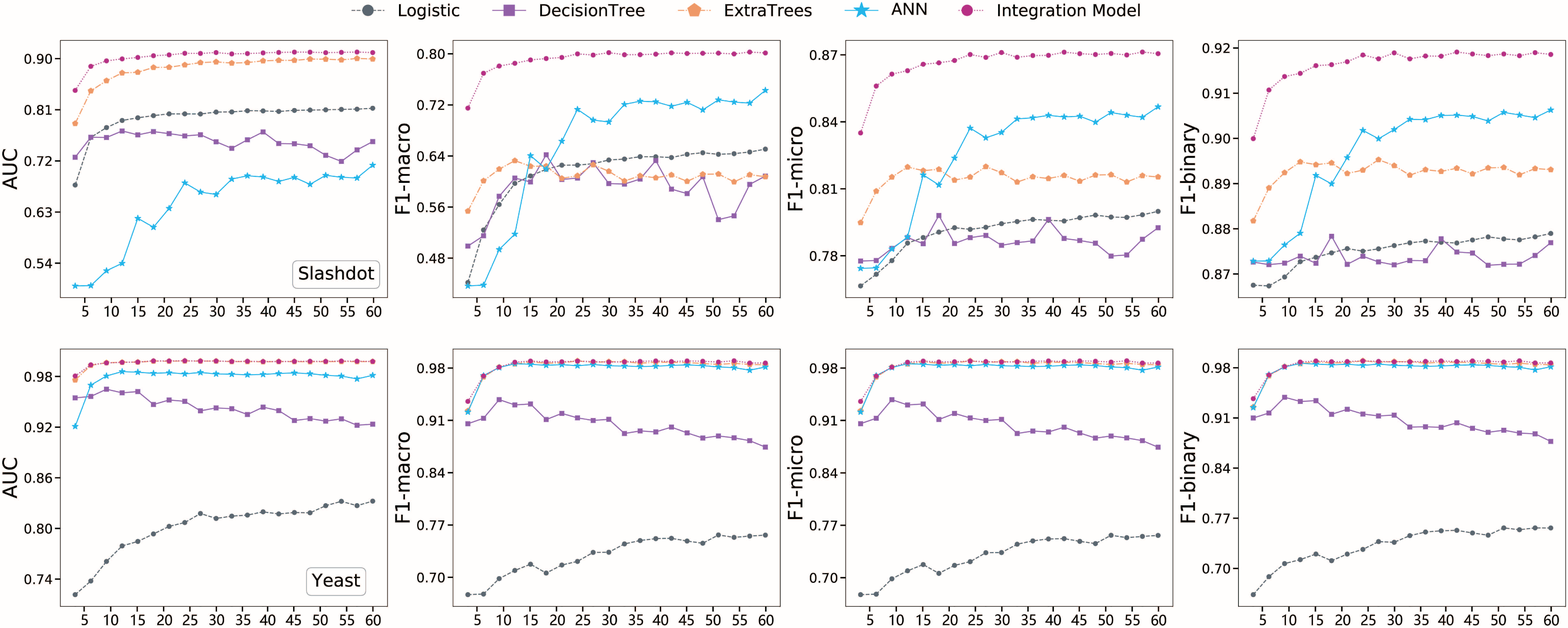}}
\caption{Impact of  node embedding dimensions on prediction, using Slashdot and Yeast, respectively,  for sign  and link prediction. Tested on two network data separately, the integration model's score was smoother when the node dimension was 10, and it had the minimum score jitter on different node dimensions. Although the artificial neural network designed according to experience performed well at link prediction, it was not satisfactory in sign prediction.}
\label{fig:dim result}
\end{figure*}
	

	\begin{table*}[!htb]
		\centering
		\caption{Comparison of the training time across various methods. ``*" Indicates That Network Feature Engineering is Deployed. }
		\begin{tabular}{cccccccccc}
			\toprule
			& SEAL & Beside & Logistic* & ANN* & Integration Model* \\
			\midrule
			Celegans & $\sim$ 2.5min &  --- & $\sim$ 0.1min & $\sim$ 0.5min & $\sim$ 2.5min \\
			PB & $\sim$ 15min & --- & $\sim$ 1min & $\sim$ 2min & $\sim$ 6min \\
			Yeast & $\sim$ 25min &  --- & $\sim$ 0.5min & $\sim$ 1.5min & $\sim$ 4min \\
			Facebook &  $\sim$ 5hour  & ---  & $\sim$ 4.5min & $\sim$ 20min & $\sim$ 30min \\
			Slashdot &  --- & $\sim$ 2hour & $\sim$ 2.5min & $\sim$ 0.5hour & $\sim$ 2hour \\
			Epinions &  ---  & $\sim$ 4hour & $\sim$ 25min & $\sim$ 2.5hour & $\sim$ 3hour \\
			Wikirfa &  --- & $\sim$ 1hour & $\sim$ 15min & $\sim$ 1hour & $\sim$ 45min \\
			\bottomrule
		\end{tabular}%
		\label{tab:time_compare}%
	\end{table*}%

\subsubsection{Sign Prediction}
	
In Table \ref{tab:result2}, the integration model performed best across all three datasets on both AUC and F1-related metrics, which shows that it is very efficient at sign prediction tasks. The most important comparison method is the \textit{Beside} model, which was proposed specifically for the sign prediction task \cite{chen2018bridge}. It performed very well on sign prediction  considering both the triangle and bridge structures across the network topology in training. In this experiment, the \textit{Beside} model significantly outperformed \textit{DW} and \textit{LINE}. However, for a network without such specific network element structures, such as with few triangles and bridges, the performance of \textit{Beside} would be very limited. The integration model used the network embedding approach to extract the network features, and such specific network structural features could be detected in the random walks in the node sequence sample in Node2Vec \cite{grover2016node2vec}. Although the improvement of the integration model over \textit{Beside} was limited, it would be more widely employed besides the networks with triangle and bridge structures. The comparison results on link prediction (Table \ref{tab:result1}) and sign prediction (Table \ref{tab:result2}) show that the integration model can   perform well at both link and sign prediction.
	
Compared to the integration model without feature engineering, the performance is greatly improved on the sign prediction comparing with the link prediction. For example, for the integration model, the AUC increased from 0.7300 to 0.9122 with feature engineering on Slashdot. Similar significant enhancement can be found with the Logistic and ANN methods. Therefore, feature engineering, which considers the combination of  node features as well as the distance between the nodes, would be critically important in sign prediction. It should be noted that the ANN method performs very well at link prediction, while its performance at sign prediction is significantly weaker than that of the integration model. Although feature engineering can improve the upper limit of performance, a suitable model is  then needed to fit the high-dimensional data.
	
\subsubsection{Model Training Time Analysis}
	
To determine the scalability of the integration model, we  tested the training time. The operating environment was a computer with 90 GB RAM and an NVIDIA Tesla P100 30G GPU. In Table \ref{tab:time_compare}, the ANN and SEAL based on the neural network model are both the duration of 50 rounds of iterative training. The symbol ``---'' means that the method was not applied to the corresponding prediction task. And the network feature dimension of the Logistic*, ANN* and the Integration Model* are set to be 128. The training times in Table \ref{tab:time_compare} were obtained by averaging  100 realizations. We compared the training time of the proposed method with SEAL and Beside, which also performed  well in  prediction tasks. The running times in training the integration model were acceptable, and slightly shorter than those of SEAL and Beside for link and sign prediction, respectively. The results show that the integration model could   obtain higher scores with reasonable  training time.

\subsubsection{Network Embedding Dimension}
	
We  studied the impact of the embedding dimension on the proposed models, with results as shown in Fig. \ref{fig:dim result}. We performed sign and link prediction using various embedding dimensions on the Slashdot and Yeast datasets, respectively. For example, dimension with 3 means that we use a 3-dimensional vector to represent the network node using the network embedding. As the embedding dimension increased, the performance of most models first increased and then stabilized. A higher-dimensional vector generally contained more information on corresponding nodes, and a higher embedding dimension led to better performance. Some models (e.g., ANN and Decision Tree) showed instability as the embedding dimension increased. The integration model increased more sharply than the other models, where it can already perform well when the embedding dimension is about 10. As the dimension of embedding increased (larger than 10), the integration model's scores were very smooth for the AUC, F1-macro, F1-micro, and F1-binary metrics, which shows that it did not need very high embedding dimensions to achieve efficient prediction.
	
\section{Conclusion}\label{sec:conclusion}
	
We proposed an integration model, which can perform the link and sign prediction in the same framework, based on network embedding. Simple and effective feature engineering was proposed to fit multi-scene networks based on the extracted original network embedding features. Network structure features were retained to the maximum extent, and adapted through the integrated classifier. Experiments on several datasets showed that the proposed model can achieve state-of-the-art or competitive performance at both link and sign prediction. The results show that feature engineering with a suitable integrated classifier can improve the performance of the model. We also found that the network embedding dimension did not need to be too high, and the structural characteristics of the network could be well expressed with a dimension of 10. This indicates that only low-dimensional network embedding vectors are required to improve the model training speed and reduce the computational complexity. Moreover, the training time of the integration model was acceptable. Although the training data involved in this work only focus on the network structure information, the node profiles can be well considered in this model framework. After the network embedding, we can add the node profiles, such as the user’s age, education and other information to generate the new node feature, which can be used to train the classifier. By the way, this work does not involve heterogeneous networks, and we will study this more complicated network for link  and symbol prediction in the future work.

\section*{Acknowledgments}
This study was partially supported by Zhejiang Provincial Natural Science Foundation of China (Grant Nos. LR18A050001 and LR18A050004), the Natural Science Foundation of China (Grant Nos. 61873080 and 61673151) and the Major Project of The National Social Science Fund of China (Grant No. 19ZDA324).
	

\bibliographystyle{unsrt}
	
\bibliography{reference}

\end{document}